\documentclass[
    ,final            
  ]
  {aipproc}

\layoutstyle{6x9}

\usepackage{amssymb}
\usepackage{amsmath}
\def\Z{{\mathbb{Z}}}

\begin{document}

\title{Virasoro  and  W-constraints for the $q$-KP hierarchy}


\author{Kelei Tian}{
  address={Department of Mathematics, University of Science and Technology of China, Hefei, 230026 Anhui,  China}
}

\author{Jingsong He}{
  address={Department of Mathematics, Ningbo University, Ningbo, 315211 Zhejiang, China. Email:
  hejingsong@nbu.edu.cn.}}

\author{ Yi
Cheng}{
  address={Department of Mathematics, University of Science and Technology of China, Hefei, 230026 Anhui,  China} 
}

\begin{abstract}
Based on the Adler-Shiota-van Moerbeke (ASvM) formula, the Virasoro
constraints and  W-constraints for the $p$-reduced $q$-deformed
Kadomtsev-Petviashvili ($q$-KP) hierarchy  are established.
\end{abstract}

\maketitle


\section{Introduction}

The origin of $q$-calculus (quantum calculus) \cite{ks,kac}  traces
back to the early 20th century. Many mathematicians have  important
works in the area of $q$-calculus and $q$-hypergeometric series. The
$q$-deformation of  classical nonlinear integrable system (also
called $q$-deformed integrable system) started in 1990's by means of
$q$-derivative $\partial_q $   instead of usual derivative $\partial
$ with respect to $x$  in the classical system. As we know, the
$q$-deformed integrable system reduces to a classical integrable
system as $q$ goes to 1.

Several $q$-deformed integrable systems have been presented, for
example, $q$-deformation of the KdV hierarchy
\cite{zhang,frenkel,kh}, $q$-Toda equation \cite{ZTAK},
$q$-Calogero-Moser equation \cite{ilive3}. Obviously, the
$q$-deformed Kadomtsev-Petviashvili ($q$-KP) hierarchy is also a
subject of intensive study in the literature
 from \cite{mas} to \cite{hetianqkp}.

The additional symmetries, string equations and Virasoro constraints
\cite{dkjm,os,dickey3,asv1,panda1,dickey2} of the classical KP
hierarchy are important since they are involved in the matrix models
of the string theory \cite{Morozov1994}. For example, there are
several new works
\cite{Morozov1998,Aratyn2003,Alexandrov,Mironov2008,tu2} on this
topic. It is quite interesting to study the analogous properties of
$q$-deformed KP hierarchy by this expanding method. In \cite{tu},
the additional symmetries of the $q$-KP hierarchy were provided.
Recently, additional symmetries and the string equations associated
with the $q$-KP hierarchy have already been reported in
\cite{tu,hetianqkp}. The negative Virasoro constraint generators
\{$L_{-n}, n\geq1$\} of the $2-$reduced $q$-KP hierarchy are also
obtained in \cite{hetianqkp} by the similar method of \cite{panda1}.

Our main purpose of this article is to give the complete Virasoro
constraint generators \{$L_{n}, n\geq-1$\}  and  W-constraints
\{$w_m, m\geq-2$\} for the $p$-reduced $q$-KP hierarchy by the
different process with negative part of Virasoro constraints given
in \cite{hetianqkp}. The  method of this paper is based on
Adler-Shiota-van Moerbeke (ASvM) formula.

This paper is organized as follows. We give a brief description of
$q$-calculus and $q$-KP hierarchy in Section 2 for reader's
convenience. The main results are stated and proved in Section 3,
which are the Virasoro constraints and W-constraints  on the $\tau$
function for the $p$-reduced $q$-KP hierarchy. Section 4 is devoted
to conclusions and discussions.

\section{$q$-calculus and $q$-KP hierarchy}

At the beginning  of the this section, Let us recall   some useful
facts of $q$-calculus \cite{kac} in the following to make this paper
be self-contained.

The Euler-Jackson $q$-difference $\partial_q$ is defined by
\begin{equation}
   \partial_q(f(x))=\frac{f(qx)-f(x)}{x(q-1)}, \qquad q\neq 1  \label{q-derivative}
   \end{equation}
and the $q$-shift operator is
$
  \theta(f(x))=f(qx).$
It is worth pointing out that  $\theta $ does not commute with
$\partial_q$, indeed, the relation $ (\partial_q
\theta^k(f))=q^k\theta^k(\partial_q f),  k\in \mathbb{Z} $ holds.
The limit of $\partial_q(f(x))$ as $q$ approaches 1 is  the ordinary
differentiation $\partial_x(f(x)) $. We denote the formal inverse of
$\partial_q$ as $\partial_q^{-1}$. The following $q$-deformed
Leibnitz rule holds
\begin{equation}
     \partial_q^n \circ f=\sum_{k\ge0}\binom{n}{k}_q\theta^{n-k}(\partial_q^kf)\partial_q^{n-k},\qquad n\in
     \Z
     \end{equation}
where the $q$-number $(n)_q=\frac{q^n-1}{q-1}$ and the $q$-binomial
is introduced as
\[
\binom{n}{0}_q=1,\qquad
\binom{n}{k}_q=\frac{(n)_q(n-1)_q\cdots(n-k+1)_q}{(1)_q(2)_q\cdots(k)_q}.
   \]
Let $(n)_q!=(n)_q (n-1)_q (n-2)_q\cdots (1)_q$, the $q$-exponent
$e_q(x)$ is defined by
$$
e_q(x)=\sum_{n=0}^{\infty}\dfrac{x^n}{(n)_q!}=\exp(\sum_{k=1}^{\infty}\frac{(1-q)^k}{k(1-q^k)}x^k).
$$

Similar to the general way of describing the classical KP hierarchy
\cite{dkjm,dickey2}, we will give a brief introduction of $q$-KP
hierarchy  and its additional symmetries based on \cite{iliev2,tu}.

The Lax operator $L$ of $q$-KP hierarchy is given by
\begin{equation}\label{qkplaxoperator}
L=\partial_q+ u_0 +
u_{-1}\partial_q^{-1}+u_{-2}\partial_q^{-2}+\cdots.
\end{equation}
where $u_i=u_i(x,t_1, t_2, t_3,\cdots,),i=0,-1,-2, -3, \cdots $. The
corresponding Lax equation of the $q$-KP hierarchy is defined as
\begin{equation}
\dfrac{\partial L}{\partial t_n}=[B_n, L], \ \ n=1, 2, 3, \cdots,
\end{equation}
here the differential part $B_n=(L^n)_+=\sum\limits_{i=0}^n
b_i\partial_q^i$ and  the integral part $L^n_-=L^n-L^n_+$.  $L$ in
eq.(\ref{qkplaxoperator}) can be generated by dressing operator
$S=1+ \sum_{k=1}^{\infty}s_k
\partial_q^{-k}$ in the following way
\begin{equation}
L=S  \partial_q  S^{-1}.
\end{equation}
Dressing operator $S$ satisfies Sato equation
\begin{equation}
\dfrac{\partial S}{\partial t_n}=-(L^n)_-S, \quad n=1,2, 3, \cdots.
\end{equation}
The $q$-wave function $w_q(x,t;z)$  and the $q$-adjoint function
$w_q^*(x,t;z)$ of $q$-KP hierarchy are given  by
$$w_q(x,t;z)=S e_q(xz) \exp({\sum  _{i=1}^{\infty}t_iz^i}),\qquad
w_q^*(x,t;z)=(S^*)^{-1}|_{x/q}e_{1/q}(-xz)\exp(-\sum_{i=1}^\infty
t_iz^i),$$ which satisfies following linear $q$-differential
equations
$$Lw_q=zw_q, \quad L^*|_{x/q}w_q^*=zw_q^*,$$
here the notation $P|_{x/t}=\sum_iP_i(x/t)t^i\partial_q^i$ is used
for a $q$-pseudo-differential operator of the form
$P=\sum_ip_i(x)\partial_q^i$, and the conjugate operation ``$*$''
for $P$ is defined by $P^*=\sum\limits_i(\partial_q^*)^ip_i(x)$ with
$\partial_q^*=-\partial_q\theta^{-1}=-\frac{1}{q}\partial_{\frac{1}{q}}$,
$(\partial_q^{-1})^*=(\partial_q^*)^{-1}=-\theta
\partial_q^{-1}$, $(PQ)^*=Q^*P^*$ for any two $q$-PDOs.

Furthermore, $w_q(x,t;z)$ and $w_q^*(x,t;z)$ of $q$-KP hierarchy can
be expressed by sole function $\tau_q(x;t)$ \cite{iliev2} as
\begin{gather}\label{qwavefunction}
  w_q=\frac{\tau_q(x;t-[z^{-1}])}{\tau_q(x;t)}
  e_q (xz)e^{\xi(t,z)}=
\frac{e_q(xz)e^{\xi(t,z)}e^{-\sum_{i=1}^\infty\frac{z^{-i}}{i}\partial_i}\tau_q}{\tau_q},
\\
  w_q^{*}=\frac{\tau_q(x;t+[z^{-1}])}{\tau_q(x;t)}e_{1/q}(-xz)
  e^{-\xi(t,z)}=
\frac{e_{1/q}(-xz)e^{-\xi(t,z)}e^{+\sum_{i=1}^\infty\frac{z^{-i}}{i}\partial_i}\tau_q}{\tau_q},
  \nonumber
\end{gather}
where $\xi(t,z)=\sum_{i=1}^{\infty}t_iz^i$ and $
[z]=\left(z,\frac{z^2}{2},\frac{z^3}{3},\ldots\right). $ The
operator $G(z)$ is  introduced as $G(z)f(t)=f(t-[z^{-1}])$, then
\begin{equation}\label{wavefunctionbygz}
w_q=\frac{G(z)\tau_q}{\tau_q}
  e_q (xz)e^{\xi(t,z)}\equiv \hat{w}_qe_q (xz)e^{\xi(t,z)}.
\end{equation}

The following Lemma shows there exist an essential correspondence
between $q$-KP hierarchy and KP hierarchy.

\noindent \textbf{Lemma 1.} \cite{iliev2} Let
$L_1=\partial+u_{-1}\partial^{-1}+u_{-2}\partial^{-2}+\cdots$, where
$\partial=\partial / \partial x$, be a solution of the classical KP
hierarchy and $\tau$ be its tau function. Then
$\tau_q(x,t)=\tau(t+[x]_q)$ is a tau function of the $q$-KP
hierarchy associated with Lax operator $L$ in
eq.(\ref{qkplaxoperator}), where
$$[x]_q=\big( x, \frac{(1-q)^2}{2(1-q^2)}x^2, \frac{(1-q)^3}{3(1-q^3)}x^3,\cdots,
\frac{(1-q)^i}{i(1-q^i)}x^i,\cdots \big).$$

Define $\Gamma_q$ and Orlov-Shulman's $M$ operator \cite{tu} for
$q$-KP hierarchy as $M= S \Gamma_q S^{-1}$ and $
\Gamma_q=\sum_{i=1}^{\infty}\Big(it_i+\dfrac{(1-q)^i}{(1-q^i)}x^i\Big)\partial
_q^{i-1}.$ The the additional flows for each pair \{$m,n$\} are
difined as follows
\begin{equation}
\dfrac{\partial S}{\partial t_{m,n}^*}=-(M^mL^n)_-S,
\end{equation}
or equivalently
\begin{equation}
\dfrac{\partial L}{\partial t_{m,n}^*}=-[(M^mL^n)_-,L], \qquad
\dfrac{\partial M}{\partial t_{m,n}^*}=-[(M^mL^n)_-,M].
\end{equation}
The additional flows ${\partial_{mn}^*}= \dfrac{\partial }{\partial
t_{m,n}^*}$  commute with the hierarchy $\partial_k=\dfrac{\partial
}{\partial t_k}$, i.e. $[\partial_{mn}^*,\partial_k]=0$ but do not
commute with each other, so they are additional symmetries [12].
$(M^mL^n)_-$ serves as the generator of the additional symmetries
along the trajectory parametrized by $t_{m,n}^*$.

 \noindent \textbf{Theorem 1.}\cite{hetianqkp} If an operator $L$ does not depend on the parameters $t_n$ and the
additional variables $t_{1,-n+1}^*$, then $L^n$ is a purely
differential operator, and  the string equations of the $q$-KP
hierarchy are given by
\begin{equation}
[L^n,\frac{1}{n}(ML^{-n+1})_+]=1, \ n=2,3,4,\cdots
\end{equation}

\section{Virasoro  and W-constraints for the  $q$-KP hierarchy}
In this section, we mainly study the Virasoro constraints and
W-constraints on $\tau$-function of the $p$-reduced $q$-KP
hierarchy. To this end, two useful vertex operators $X_q(\mu,
\lambda)$  and  $Y_q(\mu, \lambda)$ would be introduced.

The vertex operator $X_q(\mu, \lambda)$  is defined in \cite{tu} as
\begin{equation}\label{qkpxqdefine1}
X_q(\mu, \lambda)=e_q(x\mu)e_q^{-1}(x
\lambda)exp(\sum_{i=1}^{\infty}t_i(\mu
^i-\lambda^{i}))exp(-\sum_{i=1}^{\infty}\frac{\mu
^{-i}-\lambda^{-i}}{i}\partial_i).
\end{equation}
We can also denote the vertex operator  $X_q(\mu, \lambda)$ by
\begin{equation}\label{qkpxqdefine2}
X_q(\mu, \lambda)=:exp(\alpha (\lambda)-\alpha (\mu)):
\end{equation}
where the symbol :: means that we keep $t_i$ be always  left side of
$\partial_j$, and $\alpha (\lambda)=\sum \alpha _n \cdot
\frac{\lambda^{-n}}{n}$, here $ \alpha _0  = 0, \alpha _n = \partial
_n=\dfrac{\partial }{\partial t_n}$ for $n>o$, $\alpha _n =
|n|t_{|n|}+\frac{(1-q)^{|n|}}{1-q^{|n|}}x^{|n|}$ for $n<o. $

The following lemma is given without proof.

 \noindent \textbf{Lemma 2.} Taylor expansion of the $X_q(\mu,
\lambda)$ on $\mu$ at the point of $\lambda$ is
$$
X_q(\mu,
\lambda)=\sum_{m=0}^{\infty}\frac{(\mu-\lambda)^m}{m!}\sum_{n=-\infty}^{\infty}\lambda
^{-m-n}W_n^{(m)},
$$
here $\sum_{n=-\infty}^{\infty}\lambda ^{-m-n}W_n^{(m)}=\partial
_{\mu}^m X_q(\mu, \lambda)|_{\mu=\lambda}.$

The first items of $W_n^{(m)}$ are
\begin{align*}
W_n^{(o)} & =\delta _{n,0}, \\
W_n^{(1)} & =\alpha _n, \\
W_n^{(2)} & =(-n-1)\alpha _n+\sum_{i+j=n}:\alpha _i\alpha _j: \\
W_n^{(3)} & =(n+1)(n+1)\alpha _n+\sum_{i+j+k=n}:\alpha _i\alpha
_j\alpha _k:-\frac{3}{2}(n+2)\sum_{i+j=n}:\alpha _i\alpha _j:
\end{align*}

There is Adler-Shiota-van Moerbeke (ASvM) formula \cite{tu} for
$q$-KP hierarchy as
\begin{equation}
X_q(\mu, \lambda)w_q(x,t;z)=(\lambda-\mu)Y_q(\mu,
\lambda)w_q(x,t;z),
\end{equation}
where the  operator $Y_q(\mu, \lambda)$ is the  generators of
additional symmetry of $q$-KP hierarchy  as
\begin{equation}
Y_q(\mu,
\lambda)=\sum_{m=0}^{\infty}\frac{(\mu-\lambda)^m}{m!}\sum_{n=-\infty}^{\infty}\lambda
^{-m-n-1}(M^mL^{m+n})_-.
\end{equation}

ASvM formula is equivalent to the following equation
\begin{equation}\label{ASvMformulaequivalent}
\partial_{m,n+m}^{*}\tau _q=\frac{W_n^{(m+1)}(\tau _q)}{m+1}.
\end{equation}

The following theorem holds by virtue of  the ASvM formula.

 \noindent \textbf{Theorem 2.}
\begin{equation}\label{basicformula}
(\frac{W_n^{(m+1)}}{m+1}-c)\tau _q=0,\ m=0,1,2,3\cdots.
\end{equation}

 \noindent \textbf{Proof.}
Consider the condition $\partial_{m,n+m}^{*}\hat{w} _q=0$, from
eq.(\ref{wavefunctionbygz}), and denote $\tilde{\tau}_q=G(z)\tau_q$,
$$
\partial_{m,n+m}^{*}\hat{w}_q = \partial_{m,n+m}^{*}
\frac{\tilde{\tau}_q}{\tau_q}
=\frac{\tilde{\tau}_q}{\tau_q}\left(\frac{\partial_{m,n+m}^{*}\tilde{\tau}_q}{\tilde{\tau}_q}-\frac{\partial_{m,n+m}^{*}\tau_q}{\tau_q}\right)
=\hat{w}_q(G(z)-1)\frac{\partial_{m,n+m}^{*}\tau_q}{\tau_q}=0. $$
The operator $G(z)$ has the property, which is  $(G(z)-1)f(t)=0$
implies $f(t)$ is a constant, from this we can get
\begin{equation}\label{tauqc}
\frac{\partial_{m,n+m}^{*}\tau_q}{\tau_q}=c
\end{equation}
where c is constant. Combining  eq.(\ref{ASvMformulaequivalent})
with  eq.(\ref{tauqc})  finishes the proof. \hfill $\square$

Now we consider the $p$-reduced $q$-KP hierarchy, by setting
$(L^{p})_-=0$, i.e. $L^{p}=(L^{p})_+$. From Lax equation of $q$-KP
hierarchy, the $p$-reduced condition  means that $L$ is independent
on $t_{jp}$ as $\partial _{jp} L=0,  j=1,2,3,\cdots$ and $\tau _q$
is independent on $t_{jp}$ as $\partial _{jp} \tau _q=0,
j=1,2,3,\cdots$.

Based on theorem 2, the Virasoro constraints and  W-constraints for
the $p$-reduced $q$-KP hierarchy will be obtained.  Let $n=kp$ in
theorem 2 and denote
\begin{equation}
 \tilde{t}_i=t_i+\frac{(1-q)^i}{i(1-q^i)}x^i, \ i=1,2,3,\cdots.
\end{equation}

First of all, for $m=0$,  eq.(\ref{basicformula}) in theorem 2
becomes
\begin{equation}\label{wkp1}
(W_{kp}^{(1)}-c)\tau _q=0.
\end{equation}
Let $c=0$, we have that $\alpha _{kp}\tau _q=\dfrac{\partial \tau _q
}{\partial t_{kp}}=0$, it is just the condition $L^p=(L^p)_+$ for
$p$-reduced $q$-KP
hierarchy. \\

For $m=1$, it is
\begin{equation}\label{wkp2}
(\frac{W_{kp}^{(2)}}{2}-c)\tau _q=0
\end{equation}

 \noindent \textbf{Theorem 3.}  The Virasoro constraints imposed
on the tau function $\tau_q$ of the $p$-reduced $q$-KP hierarchy are
 $$ L_{k}\tau_q=0,\ \
k=-1,0,1,2,3,\cdots, $$ here
\begin{align*}
 L_{-1}&=\frac{1}{p}\sum_{\begin{array}{l}n=p+1\\n\neq0(\mbox{mod}p)\end{array}}^\infty
   n \tilde{t}_n\frac{\partial}{\partial
  \tilde{t}_{n-p}} +\frac{1}{2p}\sum_{i+j=p}ij \tilde{t}_{i}
  \tilde{t}_{j}, \\
 L_{0}&=\frac{1}{p}\sum_{\begin{array}{l}n=1\\n\neq0(\mbox{mod}p)\end{array}}^\infty
  n \tilde{t}_n\frac{\partial}{\partial
  \tilde{t}_{n}} +(\frac{p}{24}-\frac{1}{24p}), \\
L_{k}&=\frac{1}{p}\sum_{\begin{array}{l}n=1\\n\neq0(\mbox{mod}p)\end{array}}^\infty
 n \tilde{t}_n\frac{\partial}{\partial
  \tilde{t}_{n+kp}} +\frac{1}{2p}\sum_{\begin{array}{l}i+j=kp\\i,j\neq0(\mbox{mod}p)\end{array}}\hspace{-0.6cm}ij \tilde{t}_{i}
  \tilde{t}_{j}, \hspace{0.4cm} k\geq1,
\end{align*}
and $ L_{n}$ satisfy Virasoro algebra commutation relations
\begin{equation}
[L_{n},L_{m}]=(n-m)L_{(n+m)}, \  m,n=-1,0,1,2,3,\cdots.
 \end{equation}
 \noindent \textbf{Proof.} Following the results
 in eq.(\ref{wkp1}) and eq.(\ref{wkp2}), we have
\begin{equation}
(\frac{W_{kp}^{(2)}}{2}-c)\tau _q=(\frac{1}{2}\sum_{i+j=kp}:\alpha
_i\alpha _j:   -c)\tau _q=0.
\end{equation}
Define $L_k=\frac{W^{(2)}_{kp}}{p}$, let
$c=\frac{p}{24}-\frac{1}{24p}$ in $L_0$, otherwise $c=0$.  The
$p$-reduced condition $n\neq0(\mbox{mod}p)$ can be naturally added
without destroying the algebra structure,  because of
$\tilde{t}_{mp}$ is  presented together with
$\frac{\partial}{\partial
  \tilde{t}_{mp+kp}}$.

By a straightforward and tedious calculation,  the Virasoro
commutation relations
$$
[L_{n},L_{m}]=(n-m)L_{(n+m)}, \  m,n=-1,0,1,2,3,\cdots$$ can be
verified. \hfill $\square$

For $m=2$, it is
\begin{equation}
(\frac{W_{kp}^{(3)}}{3}-c)\tau _q=(\frac{1}{3}\sum_{i+j+h=kp}:\alpha
_i\alpha _j\alpha _h: -c)\tau _q=0.
\end{equation}

\noindent \textbf{Theorem 4.} Let $$w_m=
\sum_{\begin{array}{l}i+j+h=mp\\i,j,h\neq0(\mbox{mod}p)\end{array}}:\alpha
_i\alpha _j\alpha _h:,\ m\geq-2,
$$
the W-constraints  on the tau function $\tau_q$ of the $p$-reduced
$q$-KP hierarchy are
$$ w_{m}\tau_q=0,
 m\geq-2, $$
and they satisfy following algebra commutation relations
$$
[L_{n},w_m]=(2n-m)w_{n+m},n\geq-1, m\geq-2.
$$

For $m\geq3$, using the similar technique in  theorem 3 and 4, we
can deduce the higher order algebraic constrains on the tau function
$\tau_q$ of the $p$-reduced $q$-KP hierarchy.


 \noindent \textbf{Remark 1.} As we know, the $q$-deformed KP hierarchy
 reduces to the classical KP hierarchy when $q \rightarrow 1$ and
 $u_0=0$. The parameters
 $(\tilde{t}_1,\tilde{t}_2,\cdots,\tilde{t}_i,\cdots)$  tend to
 $(t_1+x,t_2,\cdots,t_i,\cdots)$ as $q \rightarrow 1$. One can
 further
 identify $t_1+x$ with $x$ in the classical KP hierarchy, i.e. $t_1+x \rightarrow
 x$. The deformation as $q$ goes to $1$ of Virasoro constraints and  W-constraints for the $p$-reduced
$q$-KP hierarchy
  are identical with the results of the classical KP hierarchy
given by L.A.Dickey \cite{dickey3} and S.Panda,
 S.Roy \cite{panda1}.

\section{Conclusions and discussions}
To summarize, we have derived the Virasoro constraints and
W-constraints of the $p$-reduced $q$-KP hierarchy in theorem 3 and 4
respectively. The results  of this paper show obviously that the
 Virasoro
constraint generators \{$L_{n}, n\geq-1$\}  and  W-constraints
\{$w_m, m\geq-2$\} for the $p$-reduced $q$-KP hierarchy  are
different with the form of the KP hierarchy. Furthermore, we also
would like to point out  the following interesting relation between
the $q$-KP hierarchy and the KP hierarchy
$$L_{n}=\hat{L}_{n}|_{t_i \rightarrow \tilde{t}_i=t_i+\frac{(1-q)^i}{i(1-q^i)}x^i}$$
and it seems to demonstrate that $q$-deformation  is a non-uniform
transformation for  coordinates $t_i \rightarrow \tilde{t}_i,$ which
is consistent with results on $\tau$ function \cite{iliev2} and the
$q$-soliton \cite{he} of the $q$-KP hierarchy. Here $\hat{L}_{n}$
\cite{dickey3,panda1} are Virasoro generators of the KP hierarchy.


\begin{theacknowledgments}
  This work is supported by the NSF of
China under Grant No. 10671187. Jingsong He is also supported by
Program for NCET under Grant No. NCET-08-0515.
\end{theacknowledgments}

\bibliographystyle{aipproc}   

\IfFileExists{\jobname.bbl}{}
 {\typeout{}
  \typeout{******************************************}
  \typeout{** Please run "bibtex \jobname" to optain}
  \typeout{** the bibliography and then re-run LaTeX}
  \typeout{** twice to fix the references!}
  \typeout{******************************************}
  \typeout{}
 }

\end{document}